\documentclass[useAMS,usenatbib]{mn2e}

\usepackage{graphicx}
\usepackage{epsfig} 
\usepackage{multirow} 
\usepackage{amssymb}
\usepackage{aas_macros}

 \title[The reversal of the SF--density relation in a massive X-ray selected galaxy cluster at $z$=1.58]
       {The reversal of the SF--density relation in a massive, X-ray selected galaxy cluster at $z$=1.58: results from \textit{Herschel}}

\author[Joana S. Santos et al.]
{J. S. Santos$^{1}$\thanks{E-mail:jsantos@arcetri.astro.it}, B. Altieri$^{2}$, I. Valtchanov$^{2}$, A. Nastasi$^{3}$, H. B\"ohringer$^{4}$, G. Cresci$^{1}$, \and
D. Elbaz$^{5}$, R. Fassbender$^{6}$, P. Rosati$^{7}$, P. Tozzi$^{1}$ and M. Verdugo$^{8}$ \\
$^{1}$ INAF - Osservatorio Astrofisico di Arcetri, Largo Enrico Fermi 5, 50125, Italy\\ 
$^{2}$ European Space Astronomy Centre (ESAC)/ESA, Villanueva de la Ca\~nada, 28691, Madrid, Spain\\ 
$^{3}$ Institut d'Astrophysique Spatiale, B\^atiment 121, Universit\'e Paris-Sud, 91405 Orsay Cedex, France\\
$^{4}$ Max-Planck-Institut f?ur extraterrestrische Physik, D-85748 Garching, Germany \\
$^{5}$ Laboratoire AIM-Paris-Saclay, CEA DSM Irfu - CNRS - Universit\'e Paris Diderot, CE-Saclay, F-91191 Gif-sur-Yvette, France\\
$^{6}$ INAF-Osservatorio Astronomico di Roma (OAR), Via Frascati 33, 00040, Monteporzio Catone, Italy \\
$^{7}$ Department of Physics and Earth Science, University of Ferrara, Via Saragat, 1 - 44122 Ferrara, Italy \\
$^{8}$ University of Vienna, Department of Astronomy, T\"urkenschanzstrasse 17, 1180, Vienna, Austria 
}

 \date{Accepted 2014 November 9.
       Received 2014 November 6.;
       in original form 2014 September 17}
 \pubyear{2014}

\begin{document}
  \maketitle

  \begin{abstract}
     Dusty, star-forming galaxies have a critical role in the formation and evolution of massive galaxies in the Universe.
     Using deep far-infrared imaging in the range 100-500$\mu$m obtained with the \textit{Herschel} telescope, we investigate
     the dust-obscured star formation in the galaxy cluster XDCP J0044.0-2033 at $z$=1.58, the most massive cluster at $z>$1.5, with a measured mass  
     $M_{200}$= 4.7$^{+1.4}_{-0.9}$$\times$10$^{14}$M$_\odot$. 
     We perform an analysis of the spectral energy distributions (SEDs) of 12 cluster members (5 spectroscopically confirmed) detected with 
     $\ge$3$\sigma$ significance in the PACS maps, all ULIRGs. The individual star formation rates (SFRs) lie in the range 155 -- 824 M$_\odot$yr$^{-1}$, 
     with dust temperatures of 24 -- 35 K.
     We measure a strikingly high amount of star formation (SF) in the cluster core, SFR ($<$ 250 kpc) $\ge$ 1875$\pm$158 M$_\odot$yr$^{-1}$, 
      4$\times$ higher than the amount of star formation in the cluster outskirts.
     This scenario is unprecedented in a galaxy cluster, showing for the first time a reversal of the SF -- density relation at $z\sim$1.6 in a \textit{massive} cluster.        
     
  \end{abstract}

  \begin{keywords} 
galaxy clusters, XDCP J0044.0-2033, Star formation, Far-infrared
  \end{keywords}

  \section{Introduction}
  
The high density environments of massive clusters are hostile to star forming galaxies, particularly in the central regions that are typically populated by 
"red \& dead" galaxies. This is especially so in the local Universe where galaxy populations in clusters have undergone several cycles of 
processes that quench star formation (SF) activity, such as e.g., mergers, tidal stripping, harassment, ram pressure stripping \citep[see][for a review] {Treu}. 
However, the evolution of the star formation rate (SFR) across different environments is a key property to understand the evolution of galaxies. 
In this respect, far-infrared (FIR) observations, that cover the peak of the spectral energy distribution (SED) of star forming galaxies, provide a 
powerful means to study the evolution of the SFR by tracing the dust obscured SF activity. 
Indeed it is now established that the infrared energy density steadily rises from $z$=0 reaching a peak between $z$=1 and 2 \citep[e.g.,][] {Magnelli}.  
The recent advent of the FIR space telescope \textit{Herschel} has enabled a leap forward in this field that is still unfolding 
 \citep[e.g.,][for recent reviews] {Lutz,Casey}. 

The SF -- density relation observed in the local Universe states that star forming galaxies prefer 
low galaxy density environments, i.e., the field relative to clusters, and the cluster outskirts relative to the core.
At $z\sim$1, a reversal of the SF-density relation has been reported in the field \citep{Elbaz07} and in the galaxy group CLG0218.3-0510 at $z$=1.62 \citep{Tran}, 
however, this effect has yet to be identified in the most massive structures at high-$z$. Since only a handful of $\ge$10$^{14}$M$_\odot$ clusters
have been detected at very high redshifts (i.e., 1.5$<z<$2), extending this investigation to the extreme regime of 
distant, massive galaxy clusters (high density) is mostly unexplored territory.

  \begin{figure*}
\includegraphics[width=14.2cm,angle=0]{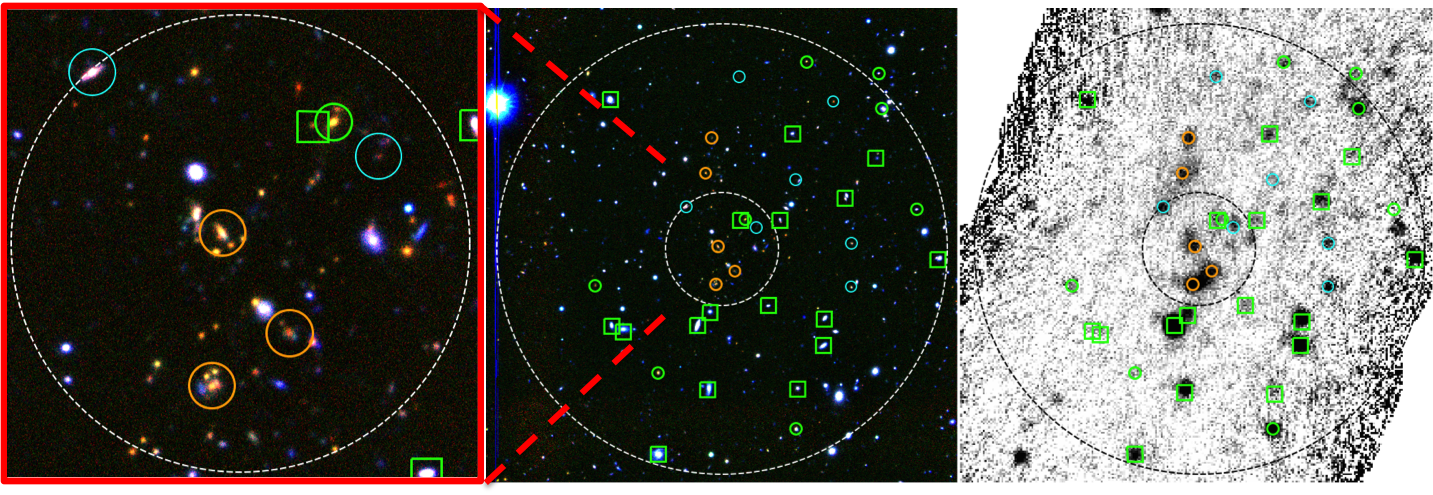} 
\vspace{-0.2cm}
 \caption{ 
 KsJI color composite of the central $\sim\! 500$ kpc (left) of the field of XDCP0044 (middle) and corresponding \textit{Herschel}/PACS 100 $\mu m$ map (right). 
 Dashed circles have radii of 30$\arcsec$ and 2$\arcmin$ centered on the cluster X-ray center. The 5 spectroscopic members with FIR 
 emission are shown in orange circles, cyan regions indicate the photometric candidates with FIR emission and green regions correspond 
 to spectroscopic (circles) and visual/photo-$z$ (squares) interlopers.}
 \label{herschelimage}
\end{figure*}

To date, the galaxy cluster XDCP J0044.0-2033 (RA: 00:44:05.2, DEC: -20:33:58, hereafter XDCP0044) at $z$=1.58 is, so far, the most massive, distant cluster 
discovered in X-rays \citep{Santos11}. This cluster was found in a serendipitous search of the XMM-Newton archive as part of the XMM-Newton 
Distant Cluster Project (XDCP), an 80 deg$^{2}$ survey aimed to discover high-redshift clusters. 
 With more than 40 galaxy clusters at 0.8$<z<$1.6, \citep[see][for a recent review] {Fassbender} the XDCP is the largest sample of X-ray selected \textit{distant} 
 clusters to date. Since the discovery of XDCP0044 we have embarked on a broad and deep observational campaign, given the uniqueness
of this system. Currently, our multi-$\lambda$ dataset includes: VI optical imaging (Subaru/Suprime), JKs near-infrared (NIR) imaging (VLT/Hawk-I), 3.6/4.5$\mu m$ 
imaging (\textit{Spitzer}/IRAC), optical (VLT/FORS2) and NIR (VLT/KMOS) spectroscopy and deep X-ray data (\textit{Chandra}).

In this Letter we present the analysis of the \textit{Herschel} \citep{Pilbratt} observations obtained for this cluster. We investigate the 
far-infrared properties of the secure cluster members as well as FIR galaxies with a photometric redshift consistent with the cluster redshift.
This study aims to provide important clues on galaxy evolutionary processes, namely the evolution of the SF -- density relation.
The cosmological parameters used throughout the paper are: $H_{0}$=70 $h$ km/s/Mpc,
$\Omega_{\Lambda}$=0.7 and $\Omega_{\rm m}$=0.3. In this cosmology, 1 Mpc at $z$=1.6 corresponds to $\sim$2$\arcmin$ on the sky.
Quoted errors are at the 1-$\sigma$ level, unless otherwise stated. 

  \section{The \textit{Herschel} data}
  
  \subsection{Observations and data reduction}
  
The \textit{Herschel} observations of XDCP0044 were carried out on December 25th, 2012, 
as a Director Discretionary Time (DDT) programme (PI Santos, 10.5 h) aimed at studying the star formation history in this massive cluster, 
following our strategy with the \textit{Herschel} Guaranteed Time programmes of high-redshift clusters and protoclusters (PI Altieri).

The PACS \citep{Poglitsch} observations at 100/160$\mu$m  (obsids = 1342257750 to 1342257759)  
were performed in mini-scan map mode, well suited for deep observations over a small field of view 
of about 2$\arcmin$  in radius.
The maps were produced using Unimap \citep{Piazzo}: a Generalized Least Square map-maker, 
that allows to reach ultimate sensitivity with no flux loss, and without iterative masking of the sources
as in the more classical masked high-pass filtering processing.
The astrometry of the PACS maps was found to be better than 1$\arcsec$, by cross-correlating
PACS sources with optical and NIR catalogues, in agreement with \textit{Herschel} absolute astrometry
accuracy (1 $\arcsec$, 1 $\sigma$).
The 3 $\sigma$ sensitivity of the maps is 1.65 (4.5) mJy  in the 100 (160)$\mu$m image.
SPIRE \citep{Griffin} observed the cluster field following an 8-point dithering pattern introduced in HerMES \citep{Oliver} in order to achieve a
more homogeneous coverage. The SPIRE maps at 250, 350 and 500 $\mu$m with nominal pixel sizes of 6$\arcsec$, 10$\arcsec$ and 14$\arcsec$ respectively, 
are dominated by the confusion noise with the estimated \textrm{rms} in the central part of the maps of 6.2, 6.5 and 7.3 mJy\footnote{
We use the median absolute deviation (MAD) as a robust measure of the noise in the presence of sources in the region. The  derived \textit{rms} corresponds to MAD 
multiplied by 1.48, which assumes a Gaussian noise distribution; this assumption is generally adequate for extragalactic SPIRE maps produced with the 
standard mapmaking pipeline.}. 

\begin{table*}
\caption{Properties of the FIR cluster members, spectroscopic (first five rows) and with photometric redshift concordant with the cluster (7 rows after line break).}  
\label{table:1}     
\small
\centering           
\begin{tabular}{llllllllll}
\hline\hline   
ID         &  RA    &        DEC  &     r$_{proj}$              &    F$_{100\mu m}$  &  F$_{160\mu m}$  & Mass  &   LIR  &  SFR   & T$_{dust}$   \\
             &           &                  &         kpc         &     mJy                      &     mJy                   &  $M_\odot$  &    $\times 10^{12} L_\odot$  & M$_\odot$yr$^{-1}$  & K   \\
\hline                            
      92   &   11.02235  &  -20.5657     &  25     &   4.5$\pm$0.6    &    8.5$\pm$1.6    &    2.0$\times 10^{11}$     &       2.6$\pm$0.3     &     452$\pm$58   & 24.0$\pm$1.0     \\
      95   &   11.02269Ê &Ê -20.5713   &     142    &  11.3$\pm$1.0   &   31.9$\pm$2.7    &    2.8$\times 10^{11}$    &       4.8$\pm$0.7     &    825$\pm$120    & 34.6$\pm$0.8     \\
      96   &   11.01973ÊÊ&Ê -20.5694   &     107    &  8.1$\pm$0.8   &    23.2$\pm$2.2   &    6.0$\times 10^{10}$    &        3.5$\pm$0.5     &    600$\pm$88     & 33.0$\pm$1.7     \\
    100   &   11.02436  &  -20.5548   &     356    &  3.6$\pm$0.6   &   16.2$\pm$1.9    &   2.4$\times 10^{11}$     &        2.9$\pm$ 0.3    &    490$\pm$49    & 27.5$\pm$1.0      \\
    101   &   11.02327  &  -20.5496   &     513    &  5.0$\pm$0.6   &     9.4$\pm$1.6    &   8.4$\times 10^{10}$      &       1.3$\pm$0.4     &    221$\pm$66     &  29.7$\pm$1.4      \\
\hline
2    &    11.01899	&      -20.54059    &   792   &  3.0$\pm$0.6   &  5.7$\pm$1.6    &   5.0$\times 10^{10}$  &   1.1$\pm$0.2   &   182$\pm$35   &   31.7$\pm$4.8    \\
4    &    11.00409	&      -20.54421    &   849   &  4.1$\pm$0.6   &  4.5$\pm$1.5    &  1.4$\times 10^{11}$   &   1.3$\pm$0.3   &   223$\pm$44   &   25.0$\pm$2.2    \\
11   &   11.02740	&      -20.55986    &   254   &  4.5$\pm$0.6   &  7.7$\pm$1.6    &  2.1$\times 10^{11}$   &   1.7$\pm$0.3   &   295$\pm$54   &   29.9$\pm$3.1    \\
13  &    11.00115     &	-20.56524    &   595   &  3.9$\pm$0.6   &  6.9$\pm$1.6    &  1.4$\times 10^{11}$   &   1.3$\pm$0.3   &   228$\pm$43   &   31.0$\pm$3.7    \\
16   &   11.00124     &	-20.57174    &   610   &  5.1$\pm$0.7    & 10.8$\pm$1.7   &  2.5$\times 10^{11}$   &   1.5$\pm$0.3   &   261$\pm$54   &   33.6$\pm$3.4    \\
33   &    11.01628    & 	-20.56292    &   193   &  1.65                 &  11.8$\pm$1.7   &  4.1$\times 10^{10}$ &   3.2$\pm$0.3   &   546$\pm$50   &   25.2$\pm$1.1    \\   
34   &    11.01006    &	-20.55585    &  468     &  1.65                  &  4.7$\pm$1.5    &  3.5$\times 10^{10}$  &   0.9$\pm$0.1   &     154$\pm$23   &    27.5$\pm$3.2     \\
\hline
\end{tabular}
\end{table*}
 
   \begin{figure}
\includegraphics[width=7.5cm,angle=0]{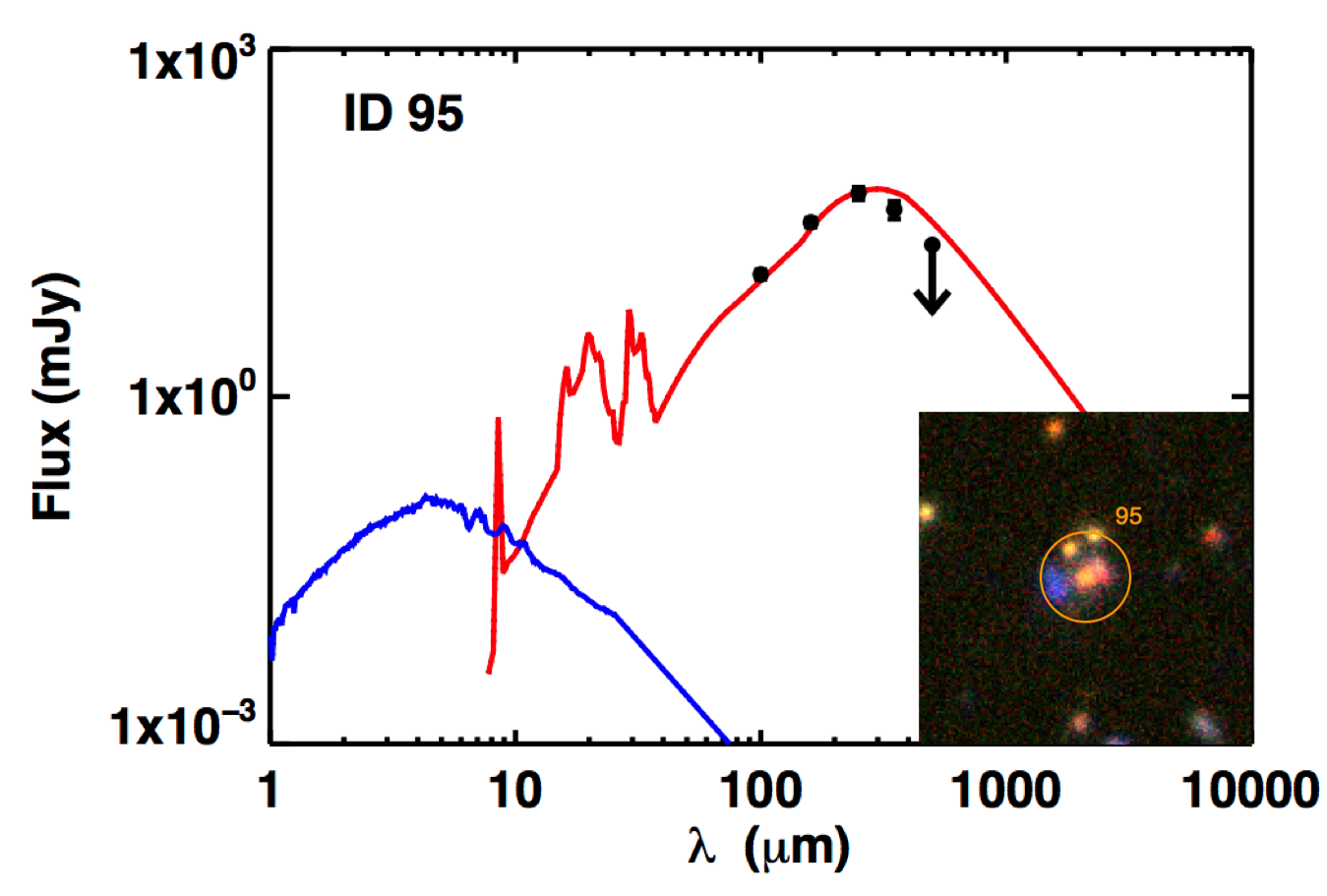} 
\vspace{-0.3cm}
 \caption{ Best-fit FIR SED (red line) of cluster member ID 95, with KsJI inset (15$\arcsec \times$15$\arcsec$). Datapoints are shown in black.}
 \label{id95}
\end{figure}
 
\subsection{Source extraction and photometry}

Our PACS maps were designed to reach very deep coverage in the cluster core, which restricted our study 
to an area within 2$\arcmin$ radius from the X-ray cluster center. This area corresponds to 1 Mpc, approximately the cluster X-ray virial radius 
\citep{Tozzi}, which makes it perfectly suited to study the global cluster star formation.

In our analysis we use a spectroscopic catalog obtained with the optical and near-IR spectroscopy campaigns using FORS2 and KMOS, respectively, 
at the VLT. Details on these observational campaigns, data reduction and results will be presented in forthcoming papers.
In summary, of the targeted 126 individual spectra, we have obtained robust redshifts for 110 galaxies. Of these, 13 are cluster members with 
$z$=[1.5701--1.6001] and the remainder 97 galaxies are interlopers. The latter are crucial to discard FIR not associated with the cluster (Fig.~\ref{herschelimage}).
Using the spectroscopic information we flag the members that are likely to be star forming based on the detection of the [OII] emission line, 
and thus have a greater probability to be detected in the \textit{Herschel} bands. Nine of the 13 members show [OII] emission.
In addition to the spectroscopic information we also use a VIJKs~3.6/4.5$\mu m$ photometric catalog covering the entire cluster field to derive 
photo-$z$s and stellar masses. 

Here we outline our procedure to reach a final catalog of FIR sources in the field of XDCP0044. 
Given the greater sensitivity of the PACS maps relative to SPIRE, we base our catalog on blind source extractions in the 100$\mu m$ 
and 160$\mu m$ maps separately, using SExtractor \citep{Bertin}. 
The Sextractor aperture photometry was validated with manual aperture photometry and corrected with the encircled energy factors 
given by \cite{Balog} in radii of 6$\arcsec$ at 100$\mu m$ and 9$\arcsec$ at 160$\mu m$. Given the difficulty to obtain reliable errors 
with standard source detection algorithms because of the correlated noise present in PACS data, we compute the photometric errors 
as the 1-$\sigma$ detection limits in each band, in addition to 7\% (calibration accuracy of the flux scale) of the source flux. 
The SPIRE source detection was performed using a simultaneous fit to all sources in the prior list \citep[XID, for more details see][] {Roseboom} 
with a prior catalogue based on the PACS detections. We do not expect problems of overdeblending (known limitation of XID) as the PACS prior 
sources of interest (see below) are separated by more than 9$\arcsec$ (FWHM/2).
We fit the three bands independently using the SPIRE point response function. 
If the fitted SPIRE flux density at the position of an input PACS source is below the $3\sigma$ sensitivity in each band we assigned the $3\sigma$ 
values as upper limits. These correspond to three times the confusion noise and are  equal to 17.4, 18.9 and 20.4 mJy at 250, 350 and 500$\mu m$, respectively.
We obtain 36 individual detections at $>$3$\sigma$ in the 160$\mu m$ map within a radius of 1 Mpc from the cluster X-ray center. Of these, 
5 are spectroscopic cluster members (4 show [OII] emission), 8 are interlopers. For the remaining 23 sources we compute photometric redshifts 
by fitting the galaxies SEDs with {\tt LePhare} \citep{Arnouts, Ilbert}, considering Bruzual \& Charlot (2003) models. 
 We assess the reliability of the photometric redshifts using our set of 13 spectroscopic redshifts in the FIR catalog and we find that, while for 4 out of 5 
confirmed spectroscopic members the photometric redshifts are a perfect match, the photo-$z$s perform more poorly for lower redshift galaxies, 
likely due to the limited optical coverage. With this in mind, we avoid selecting cluster member candidates using thresholds in photometric redshifts, and first 
discard obvious foreground galaxies using size and KsJI color information (Fig. 1, the image quality in the Ks/J/I bands is, respectively, 
0.5\arcsec/0.5\arcsec/0.7\arcsec). This leaves us with 7 galaxies with 1.0$<z_{phot}<$1.9, as 
potential cluster members detected by \textit{Herschel}. Therefore, about a third of the FIR sources in the cluster virial area are likely 
associated to the cluster (5 secure members + 7 $z_{phot}$ galaxies).

  \section{Results}
  \subsection{Far-infrared properties: LIR, SFR, Tdust}

We fit the galaxies FIR SEDs using the program {\tt LePhare}  
with Chary \& Elbaz (2001) templates, to measure individual integrated infrared luminosity, $L_{IR}$. 
The star formation rates are derived using the Kennicutt scaling relation, SFR$_{IR}$ = 1.71$\times$10$^{-10} L_{IR}/ L_\odot$ \citep{Kennicutt} that 
assumes a Salpeter initial mass function \citep{Salpeter}. 

The SFR of our sample spans the range 155 -- 824 M$_\odot$yr$^{-1}$, with an average and median of 373 and 295 M$_\odot$yr$^{-1}$, respectively.
For the candidate member galaxies the IR properties were derived with $z$=1.58, the cluster $z$.

The galaxies dust temperature ($T_{dust}$), anchored by the peak of the FIR SED, is computed with a modified black body model with an emissivity 
index $\beta$ fixed to 1.5. We find $T_{dust}$ in the range 24-35 K, typical of high-$z$ ULIRGs \citep[e.g.,][] {Hwang}.

  \subsection{sSFR -- M*}

\begin{figure*}
\includegraphics[width=7.5cm,height=5.cm,angle=0]{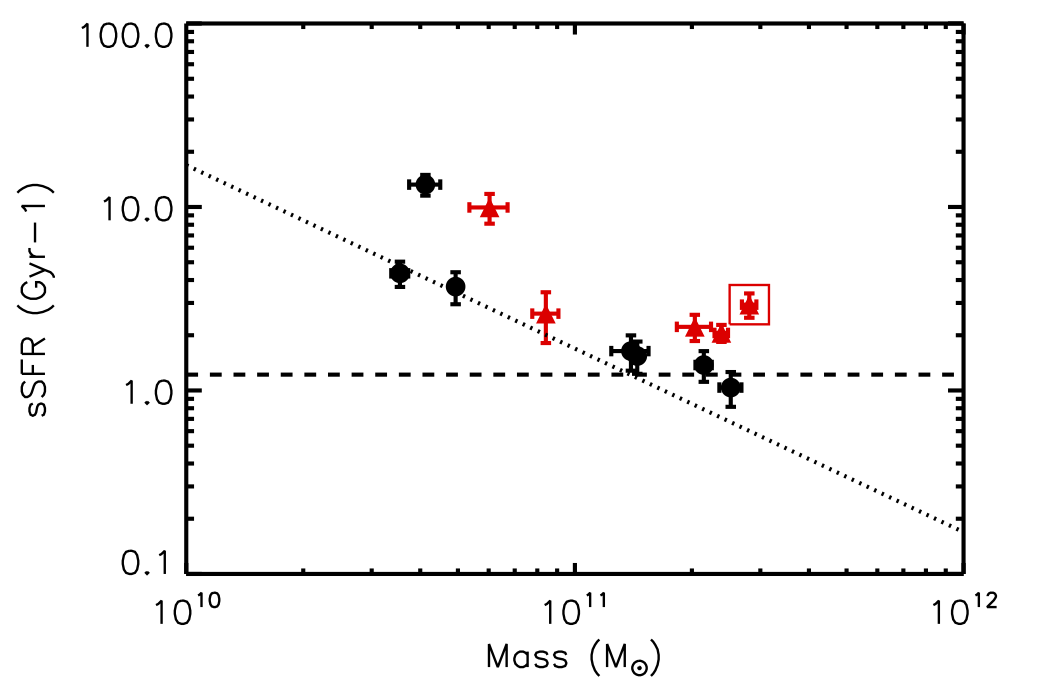} 
\includegraphics[width=7.5cm,height=5.cm,angle=0]{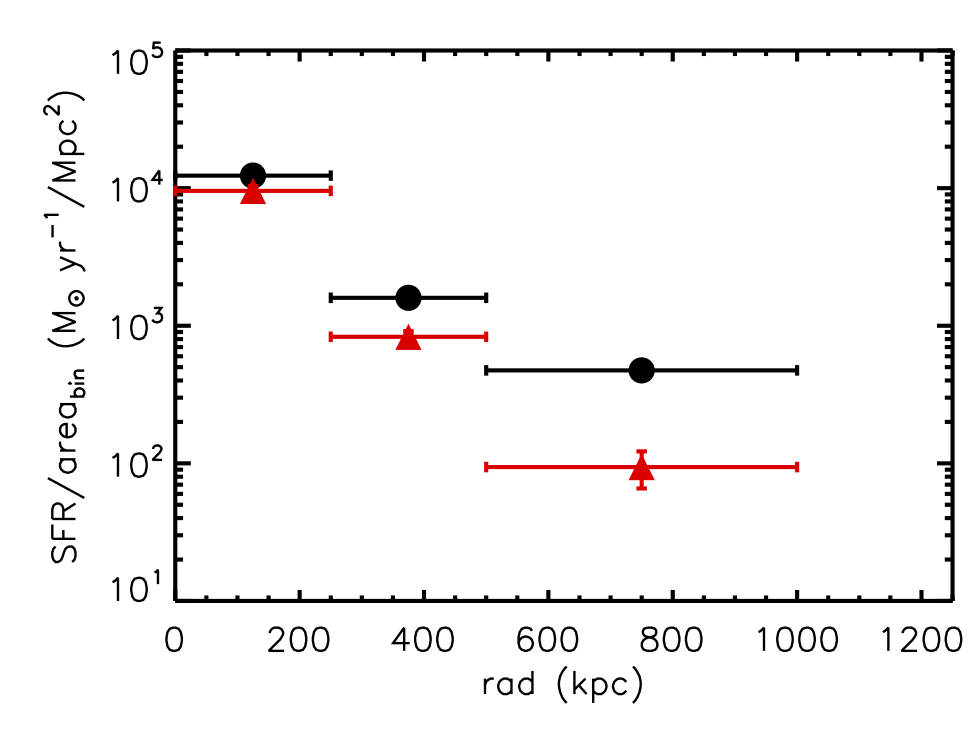} 
\vspace{-0.3cm}
\caption{(Left) sSFR vs stellar mass of the spectroscopic members (red triangles) and photometric redshift candidates (black circles) 
with \textit{Herschel} emission, in the galaxy cluster XDCP0044. The dash line indicates the expected main-sequence level of galaxies 
at the cluster redshift and the dotted line marks the 3-$\sigma$ detection limit of our \textit{Herschel} data. The square indicates galaxy ID 95. (Right) 
Differential SFR divided by bin area vs projected radius of the spec. members only (red) and adding the photo-$z$ candidates (black).  }
 \label{rad}
\end{figure*}

Stellar masses are obtained by fitting the optical through mid-infrared SED of the galaxies using again {\tt LePhare} but with Bruzual \& Charlot (2003) 
templates. In the left panel of Fig.~\ref{rad} we plot the specific SFR, sSFR, versus stellar mass  for the 12 galaxies with \textit{Herschel} emission. 
Unlike most of the \textit{Herschel} studies of distant clusters, for XDCP0044 we are able to effectively probe the main-sequence galaxy population 
\citep{Elbaz11} thanks to the unprecedented depth of our PACS maps (SFR$_{lim} \sim$165 M$_\odot$yr$^{-1}$). 

All FIR cluster galaxies have $\gtrsim$ 10$^{12} L_\odot$, placing them in the Ultra Luminous Infrared Galaxy (ULIRG) category. 
However, as we see in Fig.\ref{rad}-left panel, these galaxies populate both the main-sequence and the starburst regions of the sSFR -- $M*$ relation.
Given the small angular size and faintness of these galaxies, the available ground-based optical and NIR imaging from Suprime/Subaru and 
Hawk-I/VLT data \citep{Fassbender14} only allow for a rough evaluation of galaxy morphology. While some cluster galaxies appear to have signs of complex 
morphology, only ID 95 has clear indication of an on-going merger with a blue companion that is also a spectroscopically confirmed member (Fig.~\ref{id95}).
 The SFRs presented for this cluster are intended as lower limits given the 
  confusion noise and the large diffraction limited beam inherent to the \textit{Herschel} observations. 
  Contamination introduced by neighbors was not a major issue because most of the FIR cluster members were fairly isolated, 
  and our ample ancillary data allowed us to assess the probability of FIR emission caused by neighbors.

  \subsection{AGN contamination}
  
It is well known that the presence of AGN may affect the interpretation of the far-infrared emission as solely star formation activity.  
Two cluster members, ID 92 and 95, are associated with X-ray AGN emission detected with the high-resolution \textit{Chandra} data \citep{Tozzi}. 
In both cases the unabsorbed X-ray luminosity of these AGN is characteristic of Type II AGN. 
 For both galaxies we investigate the contribution of the AGN component to the FIR emission using the programme {\tt DecompIR} \citep{Mullaney}, 
an SED model fitting software that attempts to separate the AGN from the host star forming galaxy.
In short, the AGN component is an empirical model based on observations of moderate-luminosity local AGNs, whereas the
5 starburst models were developed to represent a typical range of SED types, with an extrapolation beyond 100$\mu m$ 
using a grey body with emissivity $\beta$ fixed to 1.5. 
For both galaxies, the best-fit model obtained with {\tt DecompIR} indicates that the host galaxy dominates the FIR emission, 
with no relevant contribution from the AGN to the total infrared luminosity, a result that is in agreement with previous \textit{Herschel} 
studies of X-ray AGNs \citep[e.g.,][] {Rosario, Santos14}.
The results obtained here are summarized in Table 1.

  \section{Reversal of the SF-density relation in XDCP0044}

In this section we compute the total SFR in XDCP 0044 and investigate the reversal of the SF--density relation.
The total cluster star formation rate ($\Sigma$SFR) is typically reported as the integrated SFR of all members 
enclosed in a circle with 1 Mpc radius: SFR($<$1 Mpc) = 4477$\pm$215 M$_\odot$yr$^{-1}$ (2587$\pm179$ considering only the 5 spectroscopic 
members), twice as much as the total SFR of CLG0218 at the same redshift, 2087$\pm$175 M$_\odot$yr$^{-1}$ \citep{Santos14}. 

The mass-normalized cluster SFR, obtained by dividing $\Sigma$SFR by the system gravitational mass, 
$\Sigma$SFR/M$_{g}$, has been widely used to quantify the evolution of the global star formation rate in clusters with redshift. 
Recently, \cite{Webb} used a representative sample with 42 clusters selected in the infrared and derived SFRs from 24$\mu m$ fluxes to
parametrize the redshift evolution of the $\Sigma$SFR/M$_{g}$ up to $z \sim$1. The authors fit a power law to the data $\propto$ (1+$z$)$^{5.4}$, indicating a rapid 
increase of the mass-normalized SFR with redshift from which we derive an expected value of $\sim$330 M$_\odot$yr$^{-1}$ /10$^{14}$M$_\odot$ at $z$=1.58. 
Similarly, \cite{Popesso} presents empirical relations for the evolution of the $\Sigma$SFR/M$_{g}$ for groups and clusters based on \textit{Herschel} data
which implies an expected value of $\sim$80 M$_\odot$yr$^{-1}$ /10$^{14}$M$_\odot$ at $z$=1.6 for massive clusters. 
The main caveat for these predictions is the limited redshift range of the samples, $z<$1.
Using the recent X-ray measurement published in \cite{Tozzi}, $M_{200}$= 4.7$^{+1.4}_{-0.9}$$\times$10$^{14}$M$_\odot$ we 
obtain $\Sigma$SFR/M$_{g}$ = 953 M$_\odot$yr$^{-1}$ /10$^{14}$M$_\odot$, a factor 3 to 12 times higher than the above mentioned empirical expectations.

 To investigate the reversal of the SF -- density relation in this cluster we measure the  
differential SFR in three radial bins (Fig.~\ref{rad}, right) normalized by the bin area, considering either only the spectroscopic members 
(red) and both spectroscopic and photo-$z$ members (black). In both cases the area normalized SFR steeply declines with 
projected cluster centric radius, by a factor 26 (all) to 100 (spec. only) from the core ($r_{1}$=250 kpc) to the outskirts ($r_{3}$=1 Mpc).
The number of galaxies per bin is [3, 1, 1]  ([4, 3, 5]) for the spectroscopic members (all), at [$<$250, 250--500, 500--1000] kpc. 
The projected integrated SFR within the core ranges from 1875$\pm$154 
(spec. members only) to 2421$\pm$166 M$_\odot$yr$^{-1}$ (all), which is extraordinary for a massive, non cool core cluster.

Using our spectroscopic sample of 13 members, we measure a fraction of star forming galaxies (f$_{SFG}$) in the core
equal to 0.43 whereas f$_{SFG}$ (outskirts) = 0.25, a tantalizing result that corroborates the reversal of the SF--density relation in this cluster.

Lastly, we compare the SFR in the core of XDCP0044 with that of the Spiderweb proto-cluster at $z$=2.16, a high-$z$ radio galaxy surrounded by 
Ly${\alpha}$ emitters that shows no sign of an intracluster medium (ICM) and is interpreted as a precursor of a massive cluster caught in a phase of 
very active star formation.
The SFR associated with the very massive central galaxy measured with \textit{Herschel} equals to 1390$\pm$150 M$_\odot$yr$^{-1}$ 
\citep{Seymour}. In contrast, XDCP044 has prominent ICM emission and appears to be in a process of assembly of what will become 
the central galaxy. The star formation in the core, distributed over 3 spectroscopic members, is 1875 M$_\odot$yr$^{-1}$. Therefore, 
these two systems appear to have different evolutionary scenarios, illustrating the pressing need for statistical samples of distant 
clusters to understand the evolutionary pathways of cluster formation.

  \section{Conclusions}

In this Letter we present deep \textit{Herschel} data of the galaxy cluster XDCP0044, aimed at measuring the 
dust-obscured star formation in one of the most massive distant clusters known.
The total SFR within the projected core area, SFR$_{spec}$($r<250 kpc$)$\ge$ 1875 M$_\odot$yr$^{-1}$, is 4$\times$ higher than the amount of star 
formation in the cluster outskirts.
This unprecedented level of SFR in XDCP0044 confirms an inversion of the SF--density relation at $z\sim$1.6, 
a result that had been found in CLG0218.3-0510 \citep{Tran, Santos14}, however, XDCP0044 is \textit{ten times} more massive than the latter. 
In contrast, XMMU J2235.3-2557 at $z\sim$1.4, 1.6$\times$ more massive than XDCP0044, has virtually no SF in the central area \citep{Santos13}. 
We are not yet in a position to draw conclusions on the SFR in the cluster population at high-$z$ because only these 3 systems 
had detailed FIR studies, and all sample studies on the evolution of the SFR in clusters stop at $z$=1.
However, our work shows that at an epoch close to the bulk of cluster formation, core galaxies in XDCP0044 are in a very active phase of star formation, 
even if the virial mass is already significantly developed. 
         
 The upcoming \textit{Hubble Space Telescope} imaging (PI Perlmutter) and grism (PI Gobat) data will enable a 
 leap forward in the study of the fainter galaxy population, placing XDCP0044 as the one of best studied clusters at a 
 lookback time of 9.5 Gyr, a crucial epoch of cluster formation.

  \section*{Acknowledgements}
  \small
  JSS and RF acknowledge funding from the European Union Seventh Framework Programme (FP7/2007-2013) under grant agreement 
  nr 267251 "Astronomy Fellowships in Italy" (AstroFIt).
  PACS has been developed by a consortium of institutes
led by MPE (Germany) and including UVIE (Austria); KUL, CSL,
IMEC (Belgium); CEA, OAMP (France); MPIA (Germany); IFSI, OAP/AOT,
OAA/CAISMI, LENS, SISSA (Italy); IAC (Spain). This development has been
supported by the funding agencies BMVIT (Austria), ESA-PRODEX (Belgium),
CEA/CNES (France), DLR (Germany), ASI (Italy), and CICYT/MCYT (Spain).
SPIRE has been developed by a consortium of institutes
led by Cardiff Univ. (UK) and including Univ. Lethbridge
(Canada); NAOC (China); CEA, LAM (France); IFSI,
Univ. Padua (Italy); IAC (Spain); Stockholm Observatory
(Sweden); Imperial College London, RAL, UCL-MSSL,
UK ATC, Univ. Sussex, STFC, UKSA (UK); Caltech, JPL, NHSC, Univ.
Colorado (USA). This development has been supported by
national funding agencies: CSA (Canada); NAOC (China);
CEA, CNES, CNRS (France); ASI (Italy); MCINN (Spain);
SNSB (Sweden); and NASA (USA). PT acknowledges funding from contract PRIN INAF 2012 ("A unique dataset to address 
the most compelling open questions about X-ray galaxy clusters")

\end{document}